\documentclass[structabstract]{aa}  
\usepackage{graphicx}
\usepackage{textcomp}
\usepackage{txfonts}
\usepackage{multirow}
\usepackage[]{natbib}
\bibpunct{(}{)}{;}{a}{}{,}
\DeclareTextSymbol{\degre}{OT1}{23}

\newcommand{\corot}{\emph{CoRoT}}

\newcommand{\flames}{\emph{FLAMES}}

\newcommand{\giraffe}{\emph{GIRAFFE}}
\newcommand{\uves}{\emph{UVES}}
\newcommand{\flamesuves}{\emph{UVES/FLAMES}}
\newcommand{\matisse}{\emph{MATISSE}}
\newcommand{\anticenter}{\emph{anticenter}}
\newcommand{\centerf}{\emph{center}}

\newcommand{\sn}{SNR}
\newcommand{\kms}{km\,s$^{-1}$}
\newcommand{\ms}{m\,s$^{-1}$}

\newcommand{\exodat}{\emph{Exo-Dat}}

\newcommand{\vsini}{$V$\,sin\,$i$}
\newcommand{\vrad}{$V_{\rm rad}$}

\newcommand{\teff}{T$_{\rm eff}$}
\newcommand{\logg}{log{\it~g}}
\newcommand{\met}{[M/H]}
\newcommand{\alf}{[$\alpha$/Fe]}
\newcommand{\ccf}{CCF}
\newcommand{\fwhm}{FWHM}

\graphicspath{{/EPSF/}{figures/}{./}}

\begin{document}

\title{Stellar characterization of \corot/Exoplanet\\fields with \matisse
\thanks{Based on observations collected with the \giraffe\ and \flamesuves\ spectrographs at the VLT/UT2 Kueyen telescope (Paranal observatory, ESO, Chile: programs 074.C-0633A \& 081.C-0413A).}$^{,}$\thanks{Full tables \ref{TabLoggf}, \ref{TabVrad}$-$\ref{tabElo}, and \ref{tabSan} are only available in electronic form at the CDS via anonymous ftp to cdsarc.u-strasbg.fr (130.79.128.5) or via {http://cdsweb.u-strasbg.fr/cgi-bin/qcat?J/A+A/523/A91}}}

\titlerunning{Stellar characterization of \corot/Exoplanet fields with \matisse}

\authorrunning{J.-C. Gazzano et al.}

  \author{J.-C.~Gazzano\inst{1,2}
         \and P.~de~Laverny\inst{2}
         \and M.~Deleuil \inst{1}
         \and A.~Recio-Blanco\inst{2}
         \and 	F.~Bouchy\inst{3,4}
				\and C.~Moutou\inst{1}
				\and	A.~Bijaoui \inst{2}
				\and	C.~Ordenovic \inst{2}
				\and	D.~Gandolfi\inst{5,6}
				\and	B.~Loeillet\inst{1}
				}
\offprints{Jean-Christophe Gazzano, \email jean-christophe.gazzano@oamp.fr }
  \institute{Laboratoire d'Astrophysique de Marseille (UMR 6110), OAMP, Universit\'e Aix-Marseille \& CNRS, 38 rue Fr\'ed\'eric Joliot Curie, 13388 Marseille cedex 13, France
        \and
        Universit\'e Nice Sophia Antipolis, CNRS (UMR 6202), Observatoire de la C\^ote d'Azur, Laboratoire Cassiop\'ee, BP 4229, 06304 Nice, France
                 \and
        Institut d'Astrophysique de Paris (UMR7095) CNRS, Universit\'e Pierre \& Marie Curie, 98bis Bd Arago, 75014 Paris, France
			\and 
			Observatoire de Haute-Provence, CNRS/OAMP, 04870 St Michel l'Observatoire, France  
			\and
			Th\"uringer Landessternwarte Tautenburg, Sternwarte 5, D-07778 Tautenburg, Germany
			\and 
			Research and Scientific Support Department, European Space Agency (ESA-ESTEC), P.O. Box 299, 2200 AG Noordwijk, The Netherlands}

\date{Received April 1, 2010; accepted July 8, 2010}

\abstract
{}
{The homogeneous spectroscopic determination of the stellar parameters is a mandatory step for transit detections from space. Knowledge of which population the planet hosting stars belong to places constraints on the formation and evolution of exoplanetary systems.}
{We used the \flames/\giraffe\ multi-fiber instrument at ESO to {spectroscopically} observe samples {of stars} in three \corot/Exoplanet fields{, namely the \emph{LRa01}, \emph{LRc01}, and \emph{SRc01} fields}, and characterize their stellar populations. We present accurate atmospheric parameters, \teff, \logg, \met, and \alf\ derived for 1\,227 stars in these fields using the \matisse~algorithm. The latter is based on the spectral synthesis methodology and automatically provides stellar parameters for large samples of observed spectra. We trained and applied this algorithm to \flames~observations covering the Mg~\textsc{i}~b spectral range. It was calibrated on reference stars and tested on spectroscopic samples from other studies in the literature. The barycentric radial velocities and an estimate of the \vsini~values were measured using cross-correlation techniques.}
{We corrected our samples {in the \emph{LRc01} and \emph{LRa01} CoRoT fields} for selection effects to characterize their FGK dwarf stars population, and compiled the first unbiased reference sample for the in-depth study of planet metallicity relationship in these \corot~fields. We conclude that the FGK dwarf population in these fields mainly exhibit solar metallicity. We show that for transiting planet finding missions, the probability of finding planets as a function of metallicity could explain the number of planets found {in the \emph{LRa01} and \emph{LRc01} \corot~fields.} This study demonstrates the potential of multi-fiber observations combined with an automated classifier such as \matisse~for massive spectral classification.}
{}

\keywords{Techniques: spectroscopic - Stars: fundamental parameters, planetary systems, general}

\maketitle
%
\section{Introduction}

Since the discovery of the first exoplanet by \cite{1995Natur.378..355M},~large-scale spectroscopic surveys for finding planets have gathered thousands of high resolution spectra. Atmospheric parameters  and chemical composition for these samples were determined by \cite{2005ApJS..159..141V} or \cite{2008A&A...487..373S}. They explored the links between the hosting stars and field stars parameters. This requires a good characterization of the fields in which planets are searched for.
The \corot\ (Convection Rotation and Transits, see \citealt{2006cosp...36.3749B} for full details) space mission has obtained light curves with very high relative photometric precision for more than $120\,000$ stars. {However, an accurate knowledge of the fundamental parameters of the stars observed by the satellite is mandatory to fully exploit this photometric database.}

We performed a first-order spectral classification of the stars in some of the \corot/Exoplanet fields using broad-band photometry obtained at the INT/La Palma. {The coordinates and magnitudes as well as these photometric spectral} types and luminosity classes of the \corot~stars are available from \exodat\ \citep{2009AJ....138..649D,2007ASPC..376..339M} and  used for the target selection and the precise placement of its photometric masks. However, this spectral classification presents some uncertainties, {e.g.,} the unknown star's reddening, chemical abundances, {and potential} binarity. Combining this photometry with intermediate resolution spectroscopy {would help} {in the determination of} the physical parameters of the stars.
\defcitealias{Loeillet2008FLAMES}{L08}	

In this context, we present a fully automated and homogeneous determination of atmospheric parameters from intermediate resolution spectra obtained with the {\flames/\giraffe\ multi-object facility.} This work is the second scientific objective of the study presented by \citet{Loeillet2008FLAMES}, hereafter \citetalias{Loeillet2008FLAMES}. In that paper, the authors demonstrated the capability of multi-fiber instruments to find planetary candidates by radial velocity techniques. We extended this study by observing a new sample of stars with the same configuration in May-June 2008, to perform a homogeneous spectral characterization of the fields observed by \corot. For that purpose, we adapted an algorithm originally developed for the spectroscopic analysis of data to be obtained with the Radial Velocity Spectrometer of the Gaia mission: MATrix Inversion for Spectral SynthEsis, \matisse\ \citep{recioblancoetal2006, 2008AIPC.1082...54B}.
We measured the effective temperature (\teff), the surface gravity (\logg), the overall metallicity (\met), and the $\alpha-$enhancement (\alf) of $1\,227$ stars, the barycentric radial velocity of $1\,534$ stars, and estimated the projected rotational velocity (\vsini) of $1\,604$ stars located in the \emph{LRa01}, \emph{LRc01}, and \emph{SRc01} fields. 
The whole data set, {observational set-ut and date,} and the {derived physical parameters} are available to the community through the \corot\ database, \exodat\footnote{{http://lamwws.oamp.fr/exodat/}\label{FootEXO}}.
We applied the relation linking the probability of finding planets with the metallicity of the host stars \citep{2007ARA&A..45..397U} to the de-biased sample.

The structure of this paper is as follows. Section~\ref{SecObs} describes the observations, the instrument setup, and {the target selection}. Section~\ref{SecData} continues with data reduction and processing. 
The automatic algorithm, its implementation, and limitations are developed in Sect.~\ref{SecMATISSE}. 
The results and a discussion can be found in Sects.~\ref{SecRes} and \ref{SecDis}, respectively.

\section{Observations\label{SecObs}}

In January 2005 and May-June 2008, we obtained 13 half-nights in visitor mode {to perform spectroscopic observations with the \flames~multi-object facility coupled with the \giraffe\ and \uves\ spectrographs (programs 074.C-0633A and 081.C- 0413A). 
The instrument is mounted on the 8.2\,m} Kueyen telescope (UT2) based at the ESO-VLT.

The {\flames/\giraffe} observations were performed {in the MEDUSA configuration, using the} HR9B spectral domain. This setup covers about 200~\AA\ {centered at $5258$~\AA\ around the Mg~\textsc{i}~b lines,} with an intermediate resolving power ($R=25\,900$) {and a CCD pixel sampling of 0.05~\AA}. This instrumental configuration was selected in 2005 for it contains many thin spectral lines leading to a good radial velocity accuracy \citep{2002SPIE.4847..184R}.
This wavelength range is also interesting for the determination of spectroscopic parameters since it contains many metallic lines that can constrain the temperature and metallicity, and some ionized spectral lines that can help to constrain the surface gravity.
{As a part of the radial velocity follow-up of \corot~exoplanet candidates}, we observed 7 additional stars, at a higher spectral resolution and across a wider wavelength range with the \flamesuves\ facility. We used the red arm of the spectrograph at a central wavelength of {5800~\AA}, covering about 2000 \AA\ with a resolving power of {about $47\,000$}.

{ 

{The 2005 campaign was dedicated to the radial velocity follow-up of selected \corot\ stars in the so-called \anticenter\ direction field (\emph{LRa01}: Long Run Anticenter 01), at Galactic coordinates $l\simeq212.2$\degre, $b\simeq-1.9$\degre. Full description of the target selection and observational strategy for this first campaign can be found in Sect.~2 of \citetalias{Loeillet2008FLAMES}. The targets observed during the 2008 campaign belong to the \emph{SRc01} (Short Run Center 01;~$l\simeq36.8$\degre, $b\simeq -1.2$\degre) and \emph{LRc01} (Long Run Center 01; $l\simeq37.7$\degre, $b\simeq-7.5$\degre) \corot\ fields, close to the Galactic \centerf\ direction. The observation strategy for the 2008 campaign was slightly different from the 2005 one since the main purpose of the \giraffe\ observations was to characterize the stellar population observed by \corot, which was only a secondary objective for the 2005 campaign.}

  \begin{table}[!t]
  \caption[]{Priority Criteria. 9 is the highest priority, 1 the lowest..\label{TabPrio}}
  \centering
  \tiny{      
        \begin{tabular}{ccccc}
           \hline
           \hline
           \noalign{\smallskip}
           Priority  &  Sp. Types & Lum. Class &  Magnitude & Variability\\
           \noalign{\smallskip}
           \hline
           \noalign{\smallskip}
           9 &  F , G , K , M &  IV , V & $r' \leq 14$ &{no}   \\
           6 &  F , G , K , M & IV , V &{ $14 \leq r' \leq 15 $} & no \\
           3 &  F , G , K , M & IV , V &  { $14 \leq r' \leq 15 $} &  yes/no  \\
           1 &  \multicolumn{2}{c}{All Other Stars} & $r' \geq 15$ & yes/no \\
           \noalign{\smallskip}
           \hline
        \end{tabular}
    }
    \tablefoot{This selection was performed using the information available in \exodat. The last column is a flag on the probability for the star to be variable \citep{2007A&A...475.1159D}.}
\end{table}

{For the 2008 campaign, each observed field was centered around the \corot\ planetary candidates observed with \uves. The \flames/\giraffe\ fiber allocation was performed taking into account the instrumental constraints (magnitude limits, fiber positions, etc.) and according to four different levels of priorities, as described in Table~\ref{TabPrio}}. 
{Using the spectral classification available in \exodat}, we gave the highest priority to solar-type {dwarf and sub-giant stars} since these were the stellar populations we wished to characterize. {High priorities were assigned to bright targets to ensure good signal-to-noise ratios (\sn s) in short exposure times. About 4-5 fibers were used to register the sky signal.} {These selection criteria were chosen to be very similar to the 2005 ones to ensure the homogeneity of the whole sample}. 
{Bad weather conditions and instrumental issues affected the second observing campaign. Out of the 20 planned configurations, we managed} to observe only 10 \flames\ fields, {as listed in the journal of the observations reported in Table~\ref{TabJour}}. 

In total, we obtained 1241 spectra of stars {located in the \emph{SRc01} and \emph{LRc01} \corot\ fields. These new spectra were analyzed together with the 772 spectra of \emph{LRa01} stars. Hereafter,} we call the pointing direction {related to the \emph{LRc01} and \emph{SRc01} fields \centerf\ and related to the \emph{LRa01} field \anticenter.}}

\begin{table}[!h]
\caption[]{Journal of the 2008 observations with the coordinates of the field centers and the adopted exposure time. \label{TabJour}}
\tiny{
\begin{tabular}{cccccc}
\hline
\hline
\noalign{\smallskip}
 \multicolumn{1}{c}{Field} &
 \multicolumn{1}{c}{Obs. Date} &
 \multicolumn{1}{c}{R.A.} &
 \multicolumn{1}{c}{Dec} &
 \multicolumn{1}{c}{Exp.T.} &
 \multicolumn{1}{c}{N} \\
 \multicolumn{1}{c}{} & 
 \multicolumn{1}{c}{} &
 \multicolumn{1}{c}{(hh:mm:ss)} &
 \multicolumn{1}{c}{(dd:mm:ss)} &
 \multicolumn{1}{c}{(Seconds)} &
 \multicolumn{1}{c}{} \\
\noalign{\smallskip}
\hline
\noalign{\smallskip}
 LRc01\_12 & 2008-06-09 
 & 19:30:14 & $-$00:00:10 & 4290 & 128 \\
 LRc01\_02 & 2008-06-09
 & 19:23:34 & +01:16:36 & 2700& 129\\
 LRc01\_04 & 2008-06-09
  & 19:25:36 & +01:26:46 & 3600& 128\\
 LRc01\_05 & 2008-05-28
 & 19:24:19 & +00:49:41 & 2401& 128\\
 LRc01\_06 & 2008-06-09
  & 19:24:17 & +00:47:09 & 3600& 126\\
 LRc01\_09 & 2008-06-12
 & 19:26:57 & +00:38:13 & 2400& 129\\
 SRc01\_01 & 2008-06-11
 & 19:04:14 & +02:02:11 & 2700& 128\\
 SRc01\_03 & 2008-06-12
  & 19:03:49 & +03:24:50 & 3600& 129\\
 SRc01\_05 & 2008-06-11
  & 19:03:00 & +02:47:01 & 2700& 129\\
 SRc01\_07 & 2008-06-11
 & 19:03:43 & +02:58:54 & 2700& 128\\
\noalign{\smallskip}
\hline\end{tabular}
}
\tablefoot{The last column gives the number of spectra registered per field. For the 2005 observations see \citetalias{Loeillet2008FLAMES}.}
\end{table}

\section{Data processing \label{SecData}}

{The 2005 frames were reduced using the \giraffe\ BaseLine Data Reduction Software \citep[girbldrs v1.12, see][]{2002SPIE.4847..184R, 2000SPIE.4008..467B}. The second epoch data were reduced using the standard ESO reduction pipeline for \giraffe~spectra (version 6.2.a2). Both pipelines apply the standard reduction processes of bias and dark subtraction, and scattered light correction. 

The spectra from the 2005 campaign were extracted with the optimum method. Those from the 2008 campaign were extracted using the standard method.
According to the standard extraction, for each localized fiber, the spectra are extracted by summing the flux of the pixels along the direction perpendicular to the dispersion axis.
The optimum extraction method uses the shape of the fiber profile to weight the flux as a function of noise \citep{1986PASP...98..609H}.
The use of optimal extraction for the second epoch data did not show a sufficient improvement to justify its use.

{For the 2005 spectra, the wavelength solution was calculated once at the beginning of the campaign} and used for the whole data-set. Instrumental drift was followed during the run with simultaneous Th-Ar lamps.
For the second epoch data, the wavelength calibration was performed using non-simultaneous Th-Ar spectra since we did not require as high a precision in radial velocity as \citetalias{Loeillet2008FLAMES}. {A sky correction was finally applied to each stellar spectrum using standard IRAF routines.}}

Combining the two campaigns, the sample of \giraffe\ data contains spectra for {1914 different \corot\ targets} with a \sn\ ratio ranging from 10 to 100. 
The analysis of such a large amount of spectra requires a completely automated and homogeneous processing of the whole sample.
To render every spectrum comparable to a reference library, we applied the following procedure.

The raw images contained a significant number of grazing cosmic rays contaminating the dispersion axis. Some spectra were polluted by these high values spikes. We corrected the data for their presence using a sigma clipping technique adjusted to the wavelength domain. This cleaning {is a mandatory step} to avoid numerical problems during the spectroscopic analysis.

We measured the barycentric radial velocity of the stars using a weighted cross-correlation of each spectrum with a numerical mask. The latter was constructed by \cite{1996A&AS..119..373B}, who identified in the solar spectrum the spectral lines relevant to radial velocity measurements.
The resulting cross-correlation function (\ccf) is fitted by a Gaussian function. As recalled by \citetalias{Loeillet2008FLAMES}, it provides the barycentric radial velocity (\vrad) and an estimate of the projected rotational velocity (\vsini). Following \cite{2002A&A...392..215S} methodology, we calibrated the relation between the \vsini\ and the broadening of the \ccf\ ($\sigma$), for the \flames/\giraffe\ instrument. 

\smallskip

\noindent {\small\vsini\ = A$\sqrt{\sigma^{2}-\sigma_{0}^{2}}$\hspace{1.4cm} where $\sigma = \frac{FWHM}{2\sqrt{2\ln{2}}}$ \normalsize \hfill(1)}

\smallskip

\begin{figure}[!h]
\includegraphics[scale=0.5]{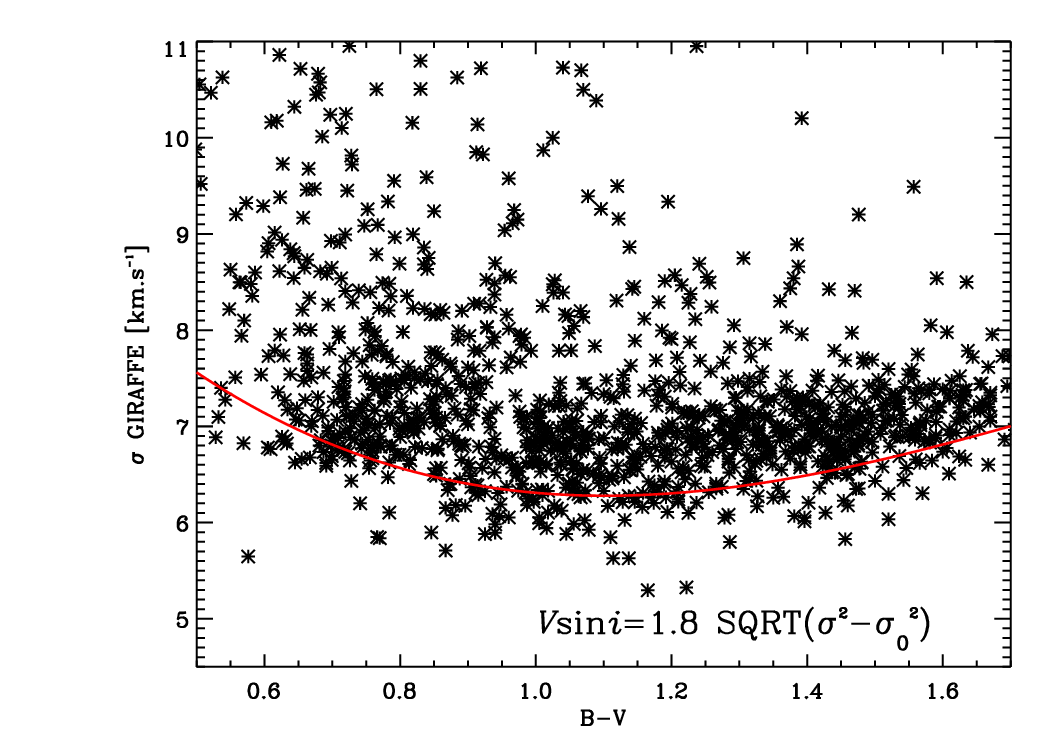}
\includegraphics[scale=0.5]{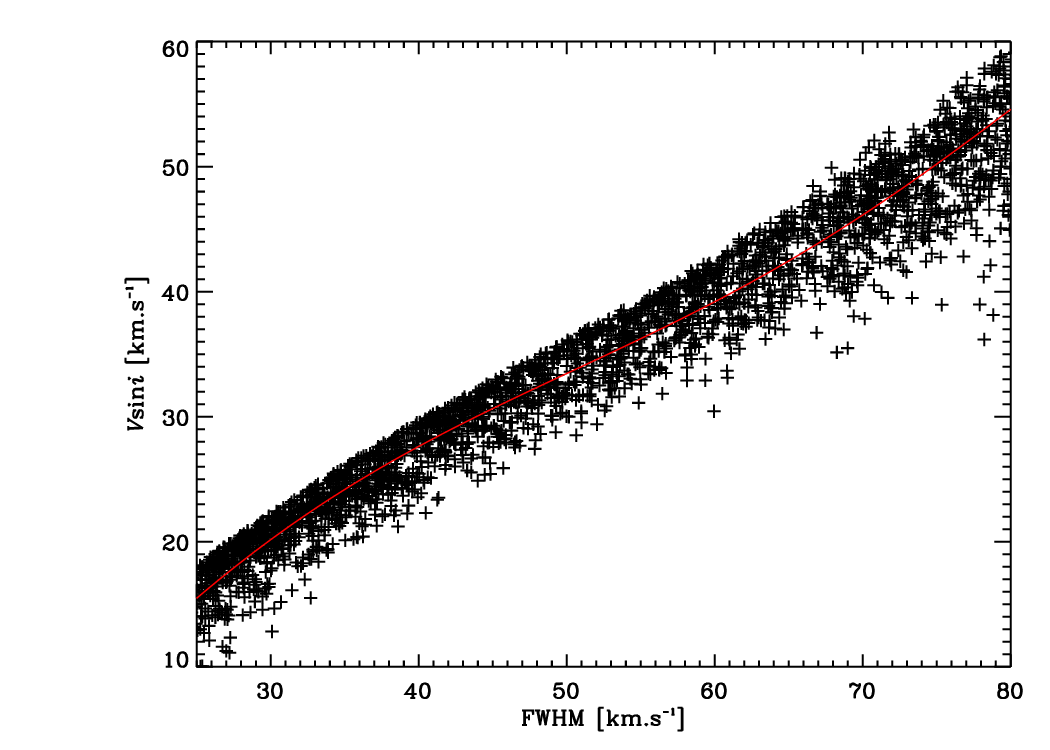}
  \caption{\textit{Top}: Calibration of the minimal broadening of the \ccf~measured as a function of the $(B-V)$ color index for slow rotators (\vsini $<$ 20 \kms). \textit{Bottom}: Correlation between the \vsini\ and the \fwhm\ of the \ccf\ for \textit{fast} rotators.}
             \label{FigVsini}
\end{figure}

\noindent For the slow rotators with typical \vsini~$<$20~\kms, we found the coefficient $A=1.8\pm0.1$ by adjusting the slope of Eq.~1 with synthetic stellar spectra. The minimal broadening {$\sigma_{0}$ in Eq.~1} as a function of the $(B-V)$ color index (see Fig.~\ref{FigVsini}, upper panel) was fitted by the relation

\smallskip

\noindent {\small $ \sigma_{0} = 10.907 - 8.361(B-V) + 4.37 6(B-V)^{2} - 0.473(B-V)^{3}$,\normalsize\hfill(2)}

\smallskip

\noindent {with the $B$ and $V$ retrieved from \exodat}. For faster rotators (\vsini~$>20$~\kms), the previous relation is no longer valid. We convolved a set of $10\,000$ interpolated synthetic spectra with different rotational broadening values ranging from 1.0 to 80.0~\kms. 
The fit of the \vsini\ as a function of the \fwhm, shown in Fig.~\ref{FigVsini} (lower panel), gives

\smallskip

\noindent {\small \vsini$\rm =-14.22~+1.04~\fwhm+0.03~\fwhm^{2}-0.001~\fwhm^{3}$.\normalsize\hfill(3)}

\smallskip

\noindent We applied Eq.~1 to the slow rotators (\vsini$< 20$~\kms) and Eq.~3 to the faster rotators. We checked that the rotational velocity calculated for stars close to 20~\kms~gave very close results with both methods. 

{The derived radial and projected rotational velocities are listed in Table~\ref{TabVrad}. A quality flag has been added for \vsini, as explained in more detail in Sect.~\ref{SecRes} and Table~\ref{TabVrad}. 
For the \emph{SRc01} field, no $V$ magnitudes are available in \exodat. 
We thus applied only Eq.~3 to derive the \vsini\ value of the stars in the \emph{SRc01} field, as clearly specified in Table~\ref{TabVrad}}.
{Taking into account the limitations of Eq.~3, we preferred to indicate a lower limit of 40~\kms\ when \vsini~$>$~40~\kms. 
{As described in the following section}, we applied \matisse\ only to the slower rotators in the sample since our tests showed that the results are not affected by the stellar rotation as long as \vsini\ is lower than $\sim11$~\kms.}

Each spectrum was then corrected for its \vrad.
The same analysis with different masks (F0, K5, M4) showed no significant improvements on the velocity precision, $\simeq0.2$~\kms, and no important mask effect on the final value. 

We estimated the position of the \textit{pseudo-continuum}, that is the apparent continuum.
For that purpose, the spectral lines were first removed by an iterative sigma-clipping method. The iterations were stopped when the dispersion of the remaining pixels reached the spectrum noise level. The position of the pseudo-continuum was then found by fitting a low order polynomial .
This first order normalized spectrum was then iteratively improved with the results from the \matisse\ analysis, as described in Sect. \ref{SecOSU}.

\begin{figure}[!h]
  \centering
  \includegraphics[scale=0.5]{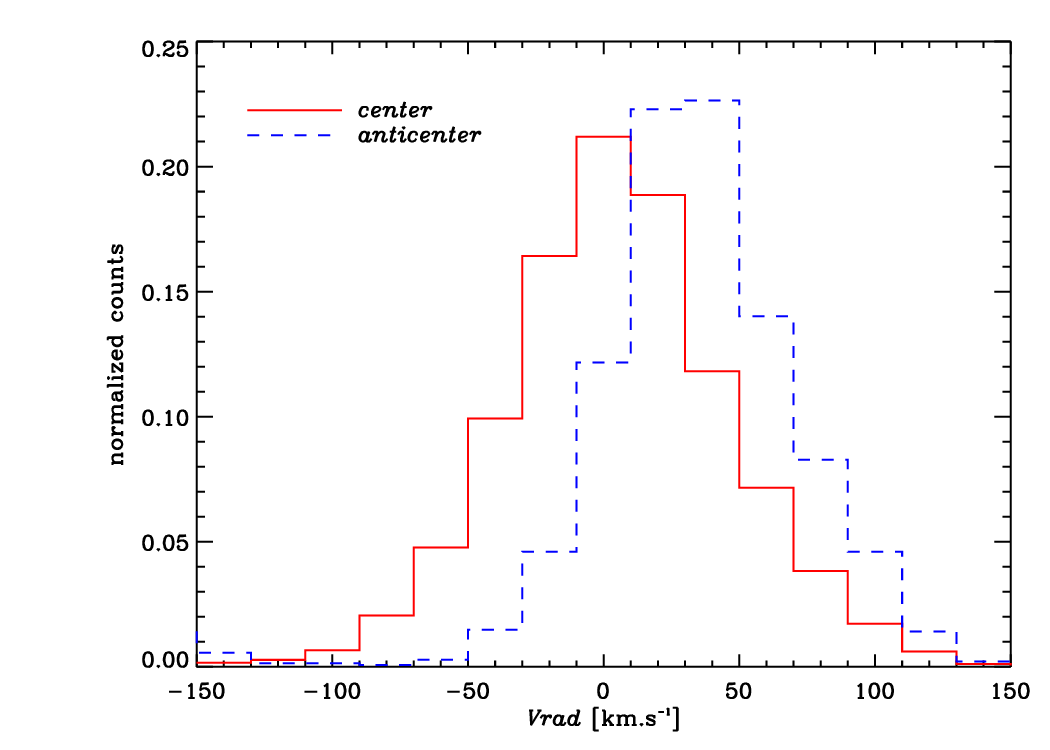}
  \includegraphics[scale=0.5]{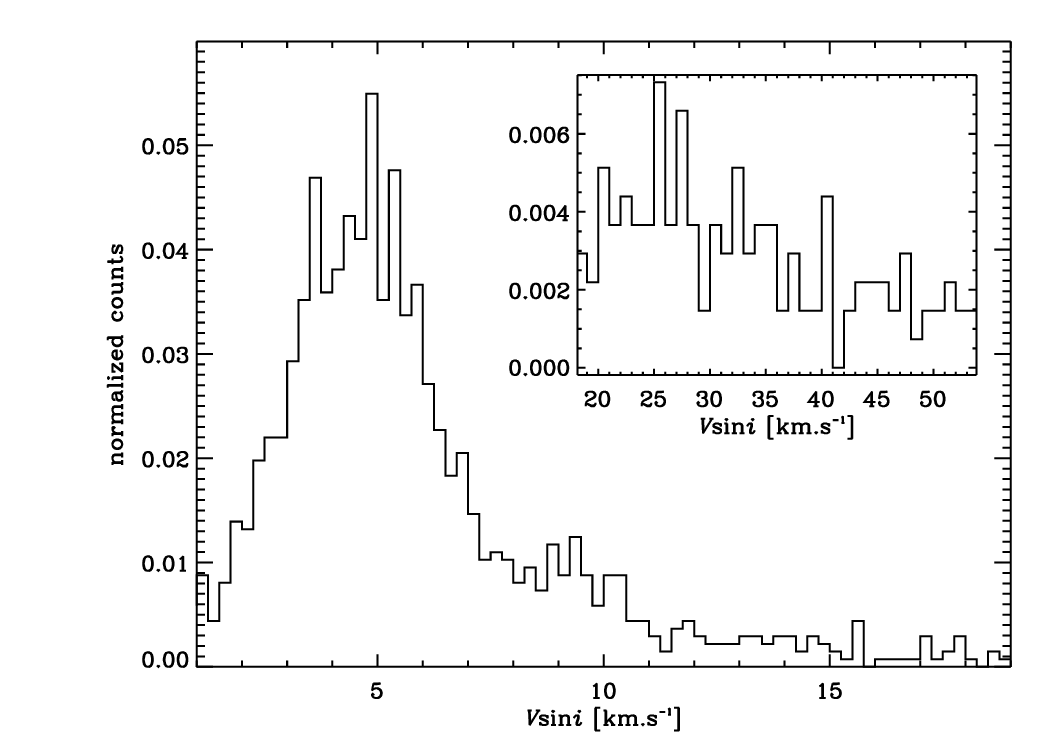}   

\caption{Kinematics results: Distribution of the  barycentric radial velocities in the two pointing directions (\textit{top}) and the projected rotational velocities (\textit{bottom}) for the whole sample.
}
             \label{FigHISTO_vsini}
\end{figure}

\section{Automatic parametrization of stellar spectra with the \matisse\ algorithm\label{SecMATISSE}}

The stellar characterization of our sample includes the determination of the following spectroscopic parameters: the effective temperature (\teff), the surface gravity (\logg), the overall metallicity (\met), and the $\alpha-$enhancement (\alf).

For that purpose, we used the \matisse\ (MAtrix Inversion for Spectral SynthEsis) algorithm, described in \cite{recioblancoetal2006}, which is connected to Local MultiLinear Regression methods. The stellar parameters are estimated by projections on relevant functions, $B(\lambda)$, derived from a multi-linear regression.
Since it was initially developed for the \emph{Gaia-RVS} instrument, it is a very efficient means of analyzing robustly and automatically large samples of stellar spectra. In the present study, we trained \matisse\ to the \flames\ instrument and in particular to the HR9B setup. 

The algorithm behavior as a function of \sn\ led us to apply an iterative inversion in the computation of the $B(\lambda)$ functions during the learning phase of \matisse\ \citep{recioblancoetal2006, 2008AIPC.1082...54B}, in the case of too noisy data.
In the following, for high \sn\ theoretical spectra, we consider the results obtained with a direct inversion method. For our observed spectra, we noted that the iterative algorithm provided more reliable results.

The next section describes the grid used for the learning phase of \matisse. {We then} describe the procedure and {estimate the errors derived by applying \matisse\ on the} stellar spectra.

\subsection{The grid of synthetic spectra \label{SecGrid}}
For the training of \matisse, a grid of theoretical spectra with the same spectral resolution and sampling as our observed data is required. For the computation of this grid, one has to keep in mind that the reliability of the atomic data is crucial when deriving accurate parameters from the stellar spectra. 
We used a list of atomic lines derived from the Vienna Atomic Line Database (VALD: \citealt{1999A&AS..138..119K}) and lists of molecular lines for the species CH, C$_2$, CN, OH, MgH, SiH, CaH, FeH, TiO, VO, and ZrO and their corresponding isotopes (kindly provided by B. Plez - see \citealt{recioblancoetal2006} for a complete description).
We calibrated the atomic line-list with observed spectra of the Sun (very high resolution Kurucz solar spectrum) and Arcturus (spectra from \citealt{2003csss...12..851H}).

The parameters used for the spectral synthesis of these stars are given in Table \ref{TabCalib} and the solar abundances are the same as those indicated by \citet{recioblancoetal2006}. For Arcturus, we used the abundances described in \cite{2000AJ....119.1239S} scaled to our solar model.

 \begin{table}[!h]
\caption[]{Atmospheric parameters used for the synthetic spectra of the stars used for checking the line-list.  \label{TabCalib}}
  \centering{      
        \begin{tabular}{cccccc}
           \hline
	    \hline
           \noalign{\smallskip}
             \multirow{2}{*}{Star}& \teff & \logg &  \met & \alf &  \multirow{2}{*}{Ref.}\\
             & (K)  & (dex) &  (dex) & (dex)\\
           \hline
           \noalign{\smallskip}
           Sun & 5777 &  4.44 & 0.0 & 0.0 &1)\\
           Arcturus &  4300 & 1.7  & $-0.6$ & 0.2 & 2) \\
           Procyon A &  6500 & 4.0  & 0.0  & 0.0  &  3)\\
           \noalign{\smallskip}
           \hline
        \end{tabular}
        }
        \tablebib{
         1) \cite{2008A&A...486..951G}, 2) \cite{2000AJ....119.1239S}, 3) \cite{2002ApJ...567..544A}}
\end{table}

For the line-list calibration, we first applied the oscillator strength modifications described in \cite{2008A&A...486..951G}, and checked that these modifications improved our fit of the solar spectrum, at our working spectral resolution and range.
We then calibrated the oscillator strengths of about 300 atomic lines at our working resolution, while verifying the coherence with the high resolution solar spectrum. 
The $\chi^2$ between the observed solar spectrum and the theoretical one was improved by a factor two. We also calibrated a few lines in the Arcturus spectrum, checking that it did not degrade the agreement on the solar spectrum. 
Finally, we checked that all these modifications also fitted the Procyon A spectrum from the \uves\ Paranal Observatory Project (\citealt{2003Msngr.114...10B} and parameters in Table \ref{TabCalib}). We preferred not to modify any oscillator strengths with this star since Procyon A parameters and furthermore its abundances are not accurately enough known. The modified atomic line data are presented in Table \ref{TabLoggf}. No modification was made to the oscillator strengths of the molecular data.

\begin{table}[!h]
\caption[]{Shortened list of the modified atomic data.\label{TabLoggf}}
\centering{      
\begin{tabular}{cccc}
\hline
\hline
\noalign{\smallskip}
 \multicolumn{1}{c}{Element} &
 \multicolumn{1}{c}{$\lambda$ (\AA)} &
 \multicolumn{1}{c}{Excitation Potential (eV)} &
 \multicolumn{1}{c}{$\log~gf$} \\
\noalign{\smallskip}
\hline
\noalign{\smallskip}

 Fe I & 5129.630 & 3.943 & -1.85\\
 Fe I & 5133.681 & 4.178 & 0.14\\
 Fe I & 5137.382 & 4.178 & -0.4\\
 Fe I & 5139.251 & 2.998 & -0.741\\
 Fe I & 5139.462 & 2.94 & -0.509\\
 Fe I & 5141.739 & 2.424 & -1.964\\
 Fe I & 5142.446 & 4.559 & -1.45\\
 Fe I & 5142.494 & 4.301 & -0.739\\
... & ...&...&...\\
           \noalign{\smallskip}
           \hline
        \end{tabular}
        }
        \tablefoot{The lines not reported here are the values from VALD. The full version of this table is available in an electronic table.}
\end{table}

From this calibrated line-list, we computed a grid of theoretical stellar spectra based on a new generation of \emph{MARCS} stellar model-atmospheres \citep{2008A&A...486..951G} with the \emph{turbospectrum} code (\citealt{1998A&A...330.1109A}, and further improvements by Plez). {The spectra were calculated assuming a plane-parallel geometry and $\xi_{t}=1$~\kms\ for stars with \logg~$\ge 3.5$~dex, and a spherical geometry, $\xi_{t}=2$~\kms, and a solar mass for star with \logg~$\le 3.0$~dex.} The calculation of the grid followed the same process as described by \cite{recioblancoetal2006}. The final library contains 11\,940 spectra and covers the spectral domain from 5\,141.70 to 5\,347.15~\AA~with a sampling of 0.07~\AA~and a resolving power of $25\,900$. The parameter ranges covered by the grid are presented in Table~\ref{TabParamsGrid}. For stars with \met~$\ge+0.0$~dex, we took \alf~$=$~0.0~dex and for stars with \met~$\le-1.0$~dex, we chose \alf~$=$~0.4~dex. Between these two metallicities, a linear relation as a function of the metallicity was computed for the \alf\ value. The \alf\ variations ($-0.4$, $-0.2$, $0.0$, $+0.2$, $+0.4$) around this law were then considered for the calculations of the spectra.

\begin{table}[!h]
\caption{Stellar atmospheric parameters of the reference grid used by \matisse.\label{TabParamsGrid}}
\centering
        \begin{tabular}{cccccc}
           \hline
	    \hline
           \noalign{\smallskip}
             & \teff & \logg & \met & \alf \\
	      &  (K)  & (dex) & (dex)& (dex)\\
           \noalign{\smallskip}
           \hline
           \noalign{\smallskip}
           Min & 3000 &  1.0 & $-3.0$ & $+0.8$ \\
           Max &  8000 & 5.5  & +1.0 & $-0.4$\\
           Steps &  200 or 250 & 0.5  & 0.25 or 0.50 & $+0.2$ \\
           \noalign{\smallskip}
           \hline
        \end{tabular}
\end{table}

To test the validity of our synthetic spectra, we applied {the whole procedure} to the spectra used for the calibration of the atomic data. {It is mandatory to verify that our procedure is self-consistent and that we do not neglect any sources of uncertainty.}
We ensured that the stellar parameters obtained for these stars are consistent with the values in Table~\ref{TabCalib}.
To test this calibration, we also processed the \giraffe\ solar atlas\footnote{\tiny{http://www.eso.org/observing/dfo/quality/GIRAFFE/pipeline/}} in exactly the same way as our \flames\ observations. The mean parameters obtained over the 129 spectra are \teff~$=5740\pm$4~K, \logg~$=4.48\pm$0.01~dex, \met~$=0.0\pm$0.004~dex, and \alf~$=0.02\pm0.002$~dex (the error bars coming from the dispersion).
These results show the reliability of this calibration and, by extension, of the results obtained in this study. {The absolute precision of our procedure was evaluated by performing a series of tests described in the following subsections.}

\subsection{Relative precision of the parameters\label{SecIE}}
To evaluate the relative precision from one star to the other, we applied \matisse\ to a large number ($10^{4}$) of interpolated synthetic spectra with randomly chosen parameters.
We added Gaussian noise to these spectra simulating \sn\ values representative of our sample: 10, 20, 50, and 100.
\begin{figure*}[!h]
  \centering
  \includegraphics[scale=1]{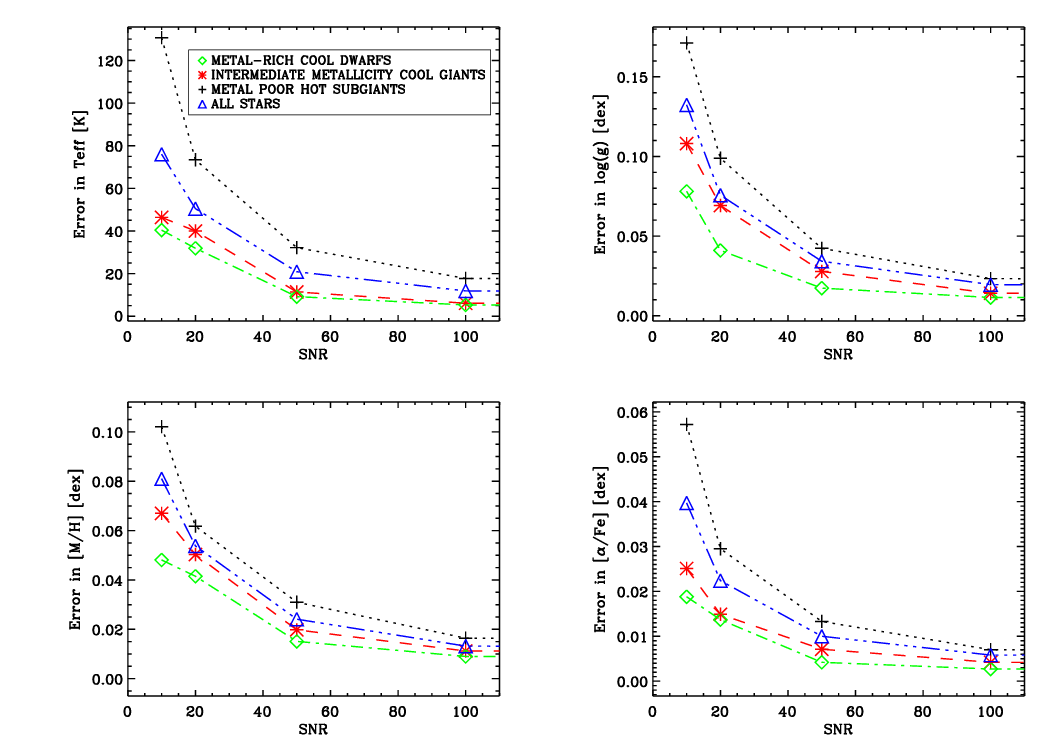}
\caption{Internal error at 70\% of the error distribution as a function of the \sn\ for cool dwarf metal-rich stars (\teff$ \le 5500$~K, \logg$\ge 4.0$~dex, $[M/H] \ge 0.0$~dex, green diamonds), cool giants with intermediate metallicity (\teff$ \le 5500$~K, \logg$\le 3.0$~dex, $0.0>[M/H] \ge -1.5$~dex red stars), hot subgiants metal poor (\teff$ > 5500$~K, $4.0>$\logg$ \ge 3.5$~dex, $[M/H] < -0.75$~dex, black crosses), and the whole sample of interpolated spectra (blue triangles).}
             \label{FigErrIntern}
\end{figure*}
\begin{table*}[!ht]
\caption{Maximum absolute value of the internal error for various proportions of the interpolated spectra at \sn=10, \sn=20 and \sn=50\label{TabPrecIn2} }
\centering{
        \begin{tabular}{cccccccccc}
           \hline
	    \hline
           \noalign{\smallskip}
            & \multicolumn{3}{c}{\sn=50} & \multicolumn{3}{c}{\sn=20} & \multicolumn{3}{c}{\sn=10} \\
             & 70\%& 80\% &90\% &70\%& 80\%&  90\%&70\%& 80\%&  90\% \\            
             \noalign{\smallskip}
           \hline
           \noalign{\smallskip}
            \multicolumn{1}{c}{\teff\ (K) } & 21 & 30 & 50 &50 &65 &98 & 76& 105  & 175\\
            \multicolumn{1}{c}{\logg\ (dex)} & 0.03 & 0.04  &  0.06 & 0.08 &0.10 &0.14 & 0.13 & 0.17 & 0.24  \\
            \multicolumn{1}{c}{\met\ (dex)} & 0.02 & 0.03 & 0.04  & 0.05 &0.07 &0.09 & 0.08 & 0.10  & 0.14 \\
           \multicolumn{1}{c}{\alf\  (dex)} & 0.01 & 0.01  & 0.02 &0.02 &0.03 &0.04 & 0.04 & 0.06 & 0.08  \\
           \noalign{\smallskip}
           \hline
        \end{tabular}
}
\end{table*}
{Figure~\ref{FigErrIntern} shows the evolution of the error at 70\% of the error distribution as a function of the \sn. This error is caused only by the dispersion in the results since the systematic uncertainty is always negligible. These dispersion errors become negligible at \sn~$>100$.
For cool metal-rich stars, which exhibit more spectral lines, more information about the stellar parameters is available, resulting systematically in lower dispersions. The analysis of the cumulative error distribution (Table \ref{TabPrecIn2} and Fig. \ref{FigErrIntern}) showed that most of the parameters are recovered with a very high accuracy.
}
\subsection{Other sources of uncertainty\label{SecOSU}}

We explored the impact on the stellar atmospheric parameters derived by \matisse\ of a potential error in the radial velocity, the normalization, and the effect of the rotational velocity. 

A large uncertainty in the radial velocity will result in a poor correction of its effects, creating an error in the determination of the various stellar parameters. Several interpolated spectra were shifted by different values of the radial velocity to evaluate these source of errors. We found that, as long as the error in the \vrad\ remains lower than 1~\kms ($\sim$10\% of the resolving power), the final uncertainty in the \matisse\ parameters is lower than the internal precision, presented in Sect. \ref{SecIE}. 
Since the mean error in the estimated radial velocity for our observed targets is about 200~\ms, we conclude that this source of uncertainty can be neglected in our study.

The rotation of the star broadens the spectral lines, and can also introduce an error in the stellar parameters. The \matisse\ algorithm was trained for non-rotating stars, which represent the vast majority of our sample (see Fig. \ref{FigHISTO_vsini}). 
Using a set of broadened synthetic spectra, we explored the effect of the stellar rotation on the photospheric parameters. 
We found that the precision of the parameters is smaller than the internal precision when \vsini~$\le$~10~\kms. 
Stars with higher rotational velocity were disregarded in this study.

Finally, we investigated the influence of the uncertainty on the shape and level of the continuum during the normalization process. For this purpose, we calculated a set of theoretical spectra at a \sn\ of 20. We modeled a departure from the true continuum by multiplying each of them by a polynomial of the second degree. Each coefficient was fixed at a level of 20\% of the error in the position, 10\% of the error in the slope, and 10\% of the error in the second order shape of the continuum. Feeding these very badly normalized spectra into \matisse, we found that the errors due to the normalization only are $\Delta$\teff=175~K, $\Delta$\logg=0.274~dex, $\Delta$\met=0.388~dex, $\Delta$\alf=0.207~dex {at most (see Sect.~\ref{SecOW}).
This study confirms that the normalization is a matter of prior importance and must be carefully taken into account. We noted that the effect is more important for high \sn\ spectra for which normalization is a critical issue.
To overcome this issue, we implemented supplementary iterations for the placement of the continuum. Each observed spectrum was divided by the synthetic spectrum calculated with the first estimate of the photospheric parameters obtained with \matisse.
The residual of the division was then cleaned from remaining spectral lines and fitted by a low order polynomial. The observed spectrum was divided by this new continuum. A new estimate of the photospheric parameters was made using this re normalized spectrum.
A few iterations were sometimes necessary before reaching a stable solution for the atmospheric parameters.
This iterative normalization ensures the minimization of this source of uncertainty that can cause large errors in the estimation of the stellar parameters.}

\subsection{Comparison with libraries of stellar atmospheric parameters\label{SecOW}}

To validate the stellar parameters derived with \matisse, we retrieved spectra from various libraries: the Elodie 3.1 archive \citep{2007astro.ph..3658P}, the S$^4$N study \citep{2004A&A...420..183A}, and the \cite{2009A&A...493..309S} work. We adapted these spectra to our spectral domain and resolution.

The Elodie library contains a quality flag for the \teff, \logg, and [Fe/H] taken from the literature. To ensure that the determination of the spectroscopic parameters are of good accuracy, we selected only the stars with high quality criteria for the \teff\ and the metallicity. These stars presented a poor \logg\ quality criterion, so we decided not to include it in our sample. 
We also selected 90 stars from the S$^{4}$N study, the remaining 29 being either fast rotators or stars for which the parameters were not determined in this study.
The surface gravity provided in the S$^{4}$N study was not derived spectroscopically. The spectroscopic method used by these authors showed a too high discrepancy with literature parameters \citep{2003ApJS..147..363A}. Hence, this parameter was derived by the algorithm described by \cite{2004A&A...420..183A} from isochrones tracks and parallaxes from the \emph{HIPPARCOS} catalogue. For this sample, we therefore compare spectroscopic gravities from \matisse\ to evolutionary ones.
In the literature, and in particular in \cite{2004A&A...420..183A}, the metallicity is inferred from the iron content ([Fe/H]). If the star appears not to have peculiar abundances, this value should be equal to the overall metallicity. For the \alf\ comparison, we used the abundances for the neutral magnesium, calcium, and silicium, determined by \cite{2004A&A...420..183A}, since these lines are present in our spectral range.
Finally, the study of \cite{2009A&A...493..309S} was interesting because it includes giant stars. We selected only these stars so as to probe the efficiency of \matisse\ for giant targets, uncovered by the two previous studies.

\begin{figure*}[!h]
  \centering
   \includegraphics[scale=1]{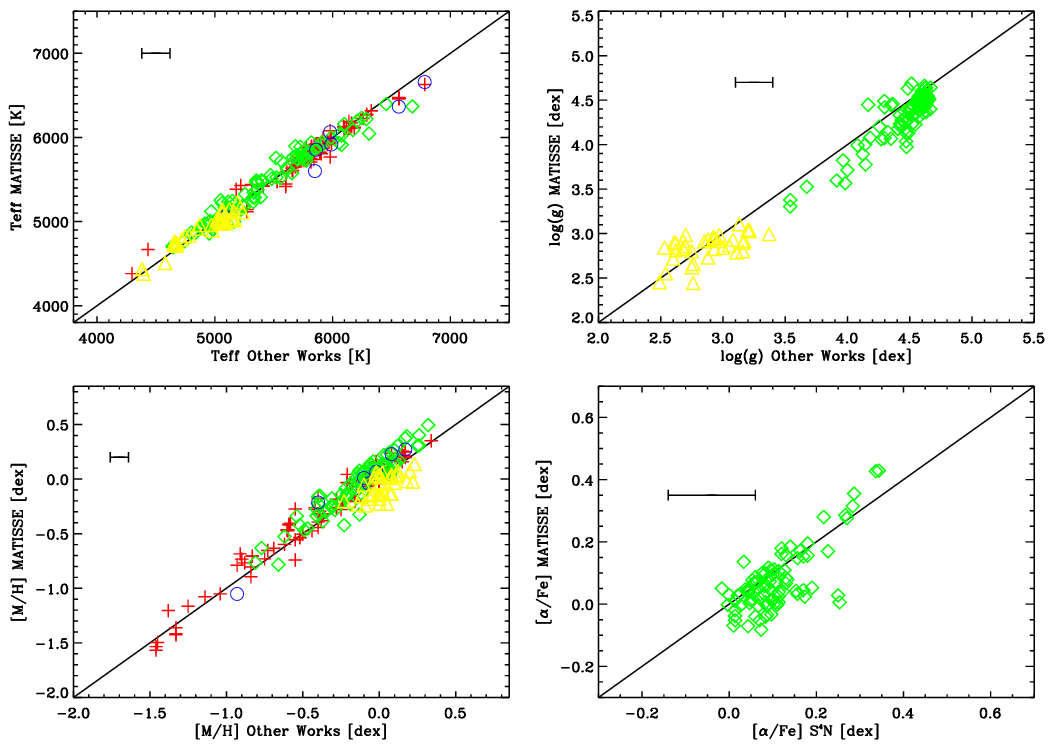}
  \caption{Comparison of the stellar atmospheric parameters derived by \matisse\ and those of the literature. Red crosses: \cite{2007astro.ph..3658P}; green diamond: \cite{2004A&A...420..183A}; blue circles: \uves\ spectra \& parameters from \cite{2007astro.ph..3658P}; yellow triangles : \cite{2009A&A...493..309S}.
  }
             \label{FigParamsParams}
\end{figure*}
In Fig.~\ref{FigParamsParams}, we compare \matisse\ spectroscopic parameters as a function of the literature ones. {The parameters derived by \matisse\ are available in electronic Tables~\ref{tabS4N}, \ref{tabElo}, and \ref{tabSan}. We find very good agreement over the four atmospheric parameters{: 70\% of these stars have difference lower than $\sim$85~K in \teff, $\sim$0.2~dex in \logg, $\sim$0.15~dex in \met, and $\sim$0.1~dex in \alf\ (see Table~\ref{TabLit}).}
This shows that for these high \sn\ observed spectra, the total error on the estimation of the parameters is at least four times higher than for interpolated synthetic spectra (\textit{internal error} negligible for \sn~$>100$ --- see Sect. \ref{SecIE} and Fig. \ref{FigErrIntern}). This illustrates that the total error in the estimation of the parameters is caused not only by the method, which exhibits an \textit{external} source of uncertainty.

We took advantage of multiple observations in our sample to evaluate the real total error on the estimation of every spectroscopic parameter. {About} fifty stars in the \centerf\ direction were observed twice and analyzed separately by the entire pipeline. The \sn s of these data range between 5 and 60. The error at 70\% of the distribution of the difference between the two determinations provides another estimate of the uncertainty. We combined this uncertainty with the \textit{internal error} and the \textit{external error} (quadratic sum) to estimate the total errors of $\sigma_{\rm T_{\rm eff}}=140$~K, $\sigma_{\log~g}=0.27$~dex, $\sigma_{\rm \met}=0.19$~dex, and $\sigma_{\rm [\alpha/Fe]}=0.09$~dex. We recall that for these samples, the relative precision from one star to another is much lower and given in Table~\ref{TabPrecIn2}.

\begin{table}[!t]
\caption{Uncertainty at 70\% of the error distribution on the measurement of stellar parameters with spectra from the literature.\label{TabLit}}
\tiny
\centering
\begin{tabular}{cccccc}
\hline
 \hline
 \noalign{\smallskip}
Nb. Spec. & \teff & \logg & \met &\alf & Ref\\
         &   (K) &   (dex)   & (dex)   &   (dex)      &    \\

            \noalign{\smallskip}
          \hline
           \noalign{\smallskip}
118 & 61 & -  & 0.11 & - & 1) \\
90 & 84 & 0.202  & 0.149 & 0.079 & 2) \\
39 & 75 & 0.184 & 0.129 & - & 3) \\
            \noalign{\smallskip}
          \hline
\end{tabular}
\tablebib{1) \cite{2007astro.ph..3658P}, 2) \cite{2004A&A...420..183A}, 3) \cite{2009A&A...493..309S} (see also Fig. \ref{FigParamsParams})}
\end{table}

\section{Results\label{SecRes}}

Table~\ref{TabBilan} reports a brief summary of the numbers of \corot\ stars presented in our study. It lists the number of observed targets, the number of stars for which we derived the physical parameters with \matisse, the number of detected double-lined spectroscopic binaries (SB2), and the number of stars with estimated radial and projected rotational velocities.  
For the targets observed in the \emph{LRa01}, we used the radial velocities as reported in the study performed by \citetalias{Loeillet2008FLAMES}. 
Most of the targets presenting no \ccf\ in the \centerf\ direction were missed due to the poor quality of the input astrometry in the \emph{SRc01} field.

\begin{figure}[!h]
  \centering
  \includegraphics[scale=0.5]{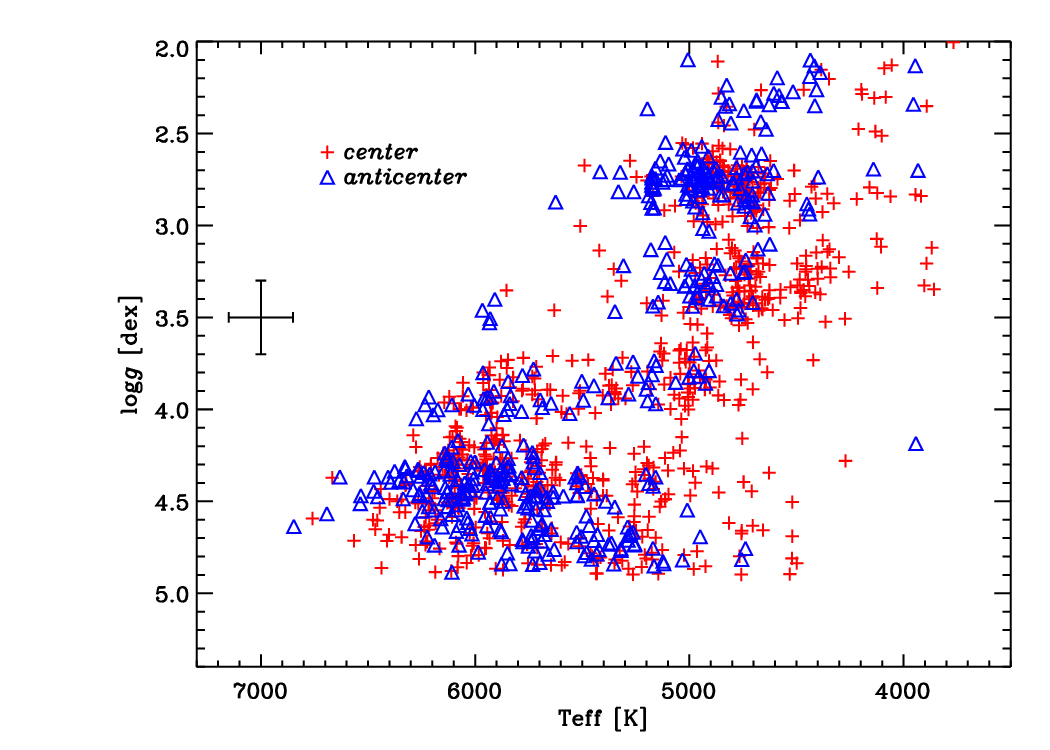}
  \caption{Hertzsprung-Russell diagram with the \teff\ and \logg\ derived by \matisse. The red crosses are stars from the \emph{LRc01} and the blue triangles from the \emph{LRa01} field.
}
             \label{FigTGM}
\end{figure}

The barycentric radial velocities, presented in Table~\ref{TabVrad}, were measured for 77\% of the stars in the \centerf\ direction and 85\% stars in the \anticenter\ direction. {We only measured precise radial velocities for stars presenting a \ccf\ with a reasonable contrast and not dominated by strong rotational broadening. The \vsini\ value was measured for $1\,604$ stars in our sample (Table \ref{TabVrad}), including objects for which \vrad\ had not been well determined\footnote{In this case, the Gaussian can still be fitted on the wings of the \ccf.}. As mentioned in Sect.~\ref{SecData}, the \vsini\ estimations have a quality flag, reported as an exponent of the value in Table~\ref{TabVrad}. Its value is set to be 1 if the \vsini\ estimate is not affected by noise or a second component in the \ccf, 2 if the target is a SB2 and the \vsini\ is related to the main component of the \ccf, and 3 if the contrast and shape of the \ccf\ are insufficient to assure a proper estimate of its parameters}. As shown in Fig.~\ref{FigHISTO_vsini}, most of the stars are slow rotators (\vsini~$<10$~\kms\ for 75\%) as expected for late-type {field} stars \citep{2008oasp.book.....G}.
In both samples, we excluded from the \matisse\ analysis the stars {exhibiting a \ccf\ with a \fwhm\ greater than 20~\kms, SB2, and stars for which no accurate radial velocity was measured because of the bad quality of the \ccf}.

\begin{table}[!b]
\caption[]{Number of targets according to the kinematics analysis.\label{TabBilan}}
  \centering
  \small{      
        \begin{tabular}{lcccc}
           \hline
           \hline
           \noalign{\smallskip}
        Number of     &  \emph{LRc01}  &  \emph{SRc01} &  \emph{LRa01} &  Total\\
                    \noalign{\smallskip}
           \hline
           \noalign{\smallskip}
	        Stars observed & 689 & 484 & 741 &  1\,914\\
		Stars analyzed by \matisse\ & 555 & 215 & 457 & 1\,227  \\
		SB2 detected & 17 & 7 & 17 & 41 \\
		\vrad\ determinations & 630 & 270 & 634 & $1\,534$ \\
		\vsini\ estimates & 658 & 268+19 & 659 & 1\,604\\
           \noalign{\smallskip}
           \hline
        \end{tabular}
    }\tablefoot{For the slow rotators in the \emph{SRc01} field, we could not apply Eq.~1 from Sect.~\ref{SecData} since no $V$ magnitude is available in \exodat. We applied Eq.~3 for these 268 targets.}
\end{table}

The \matisse\ spectral analysis, described in the previous section, was carried out on a total of $1\,227$ targets that is 65\% of the initial number of stars. The results of this analysis are presented in Table~\ref{TabVrad}.
{Combining the \teff\ and \logg\ values}, we compiled the Hertzsprung-Russell diagram presented in Fig.~\ref{FigTGM}. We observed main-sequence solar-type and cooler giant stars in each of the \corot/Exoplanet fields.

We used the \logg\ derived by \matisse\ to differentiate giants, \logg~$<3.1$~dex, and dwarfs, \logg~$>3.8$~dex. This criterion is based on the \cite{1981Ap&SS..80..353S} tables. 
The metallicity distribution for the giants and dwarfs are presented in Fig.~\ref{FigHISTOsM}. It is similar in the two pointing directions for the dwarfs. For the giants, the maxima of the metallicity distributions are apart one from the other of a few tenth of dex. This could be explained by the radial metallicity gradient generally observed in the Galactic disk (\citealt{2009A&A...504...81P} and references therein).
\begin{figure}[!h]
  \centering
\includegraphics[scale=0.5]{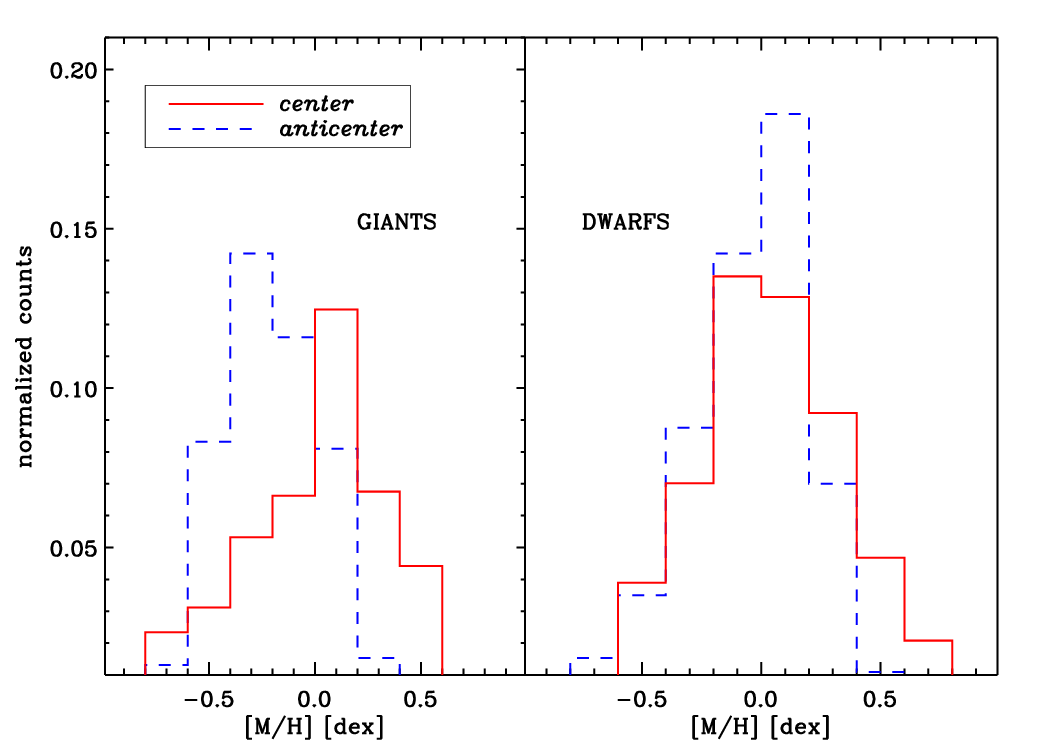}
     \caption{Distribution of the overall metallicity in the two pointing directions for giant (left) and dwarf (right) stars, normalized to the number of stars in each field. 
             \label{FigHISTOsM}
  }
\end{figure}
We compare in~Fig. \ref{FigCOLMAG} the dwarf-giant dichotomous separation based on the \logg\ derived by \matisse\ with the color-magnitude diagram \emph{r'} versus \emph{(r'$-$J)}. According to this color-magnitude diagram, the stars with an intermediate \logg\ value ([3.1$-$3.8]) could be considered as giants. We note that the dwarf-giant dichotomy used for this first order classification does not include the uncertainty in the \logg.
In the following, we include these stars in the giant sample.
{This figure also shows how the spectroscopic classification easily allows us to distinguish the dwarf and giant populations, this issue being one of the most constraining limitation of the photometric classification, especially for faint targets}.\begin{figure*}[!h]
  \centering
  \includegraphics[scale=1]{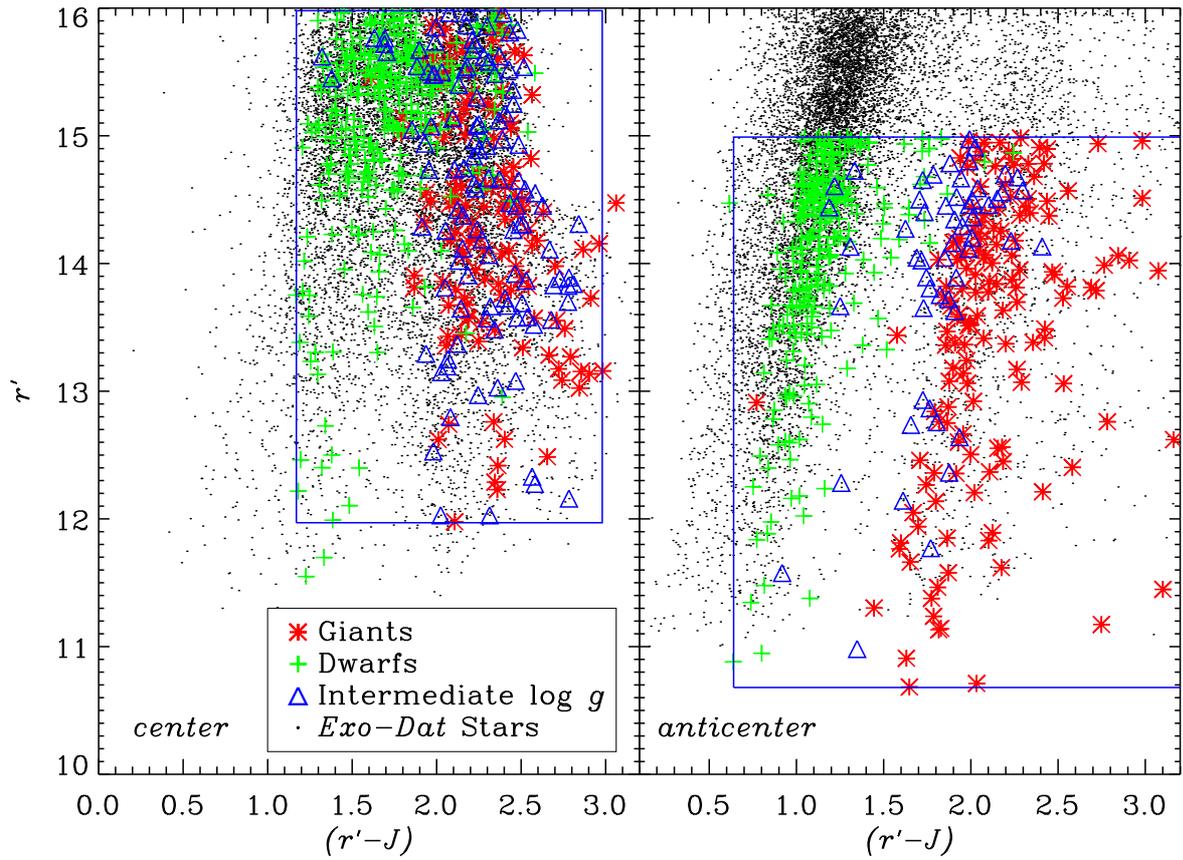}
\caption{{Color magnitude diagrams for the giants (red stars) and the dwarfs (green crosses), as well as the stars with an intermediate \logg\ (blue triangles). All the stars in the \exodat\ (black dots) are over~plotted in the diagram}. The dichotomy giant-dwarf comes from the \matisse\ parameters. The blue boxes correspond to the limits in color and magnitude for the de-biasing of our sample.
\label{FigCOLMAG} }
\end{figure*}

We find a good agreement between these simple photometric and spectroscopic classifications for the majority of our sample. However, a few stars are misclassified in both pointing directions. In the \centerf\ field, the outliers could correspond to reddened dwarfs. In the \anticenter\ direction, the obvious misclassification is due to the limits of the grid of the theoretical spectra (see Sect.~\ref{SecGrid}). {This object is a late-A metal poor star} (\met$\simeq-0.7$~dex) that exhibits very few lines in the spectral range, resulting in a very difficult classification. We found that $\sim 1$\% of the stars are misclassified.
\begin{figure*}[!h]
  \centering
  \includegraphics[scale=0.5]{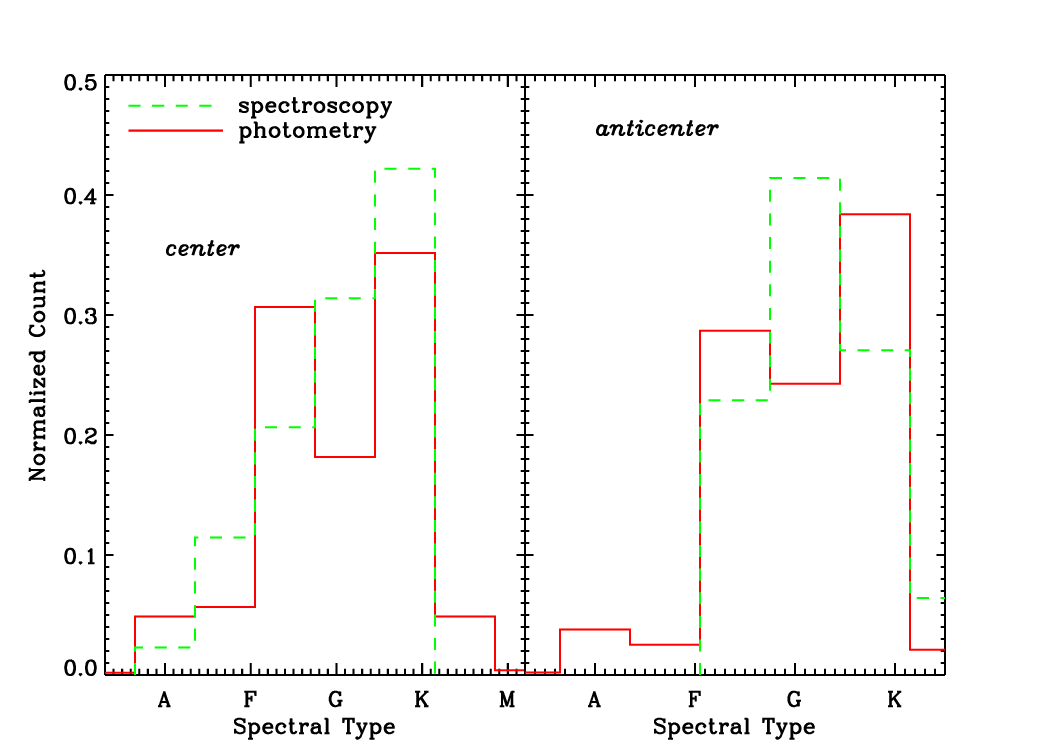}
  \includegraphics[scale=0.5]{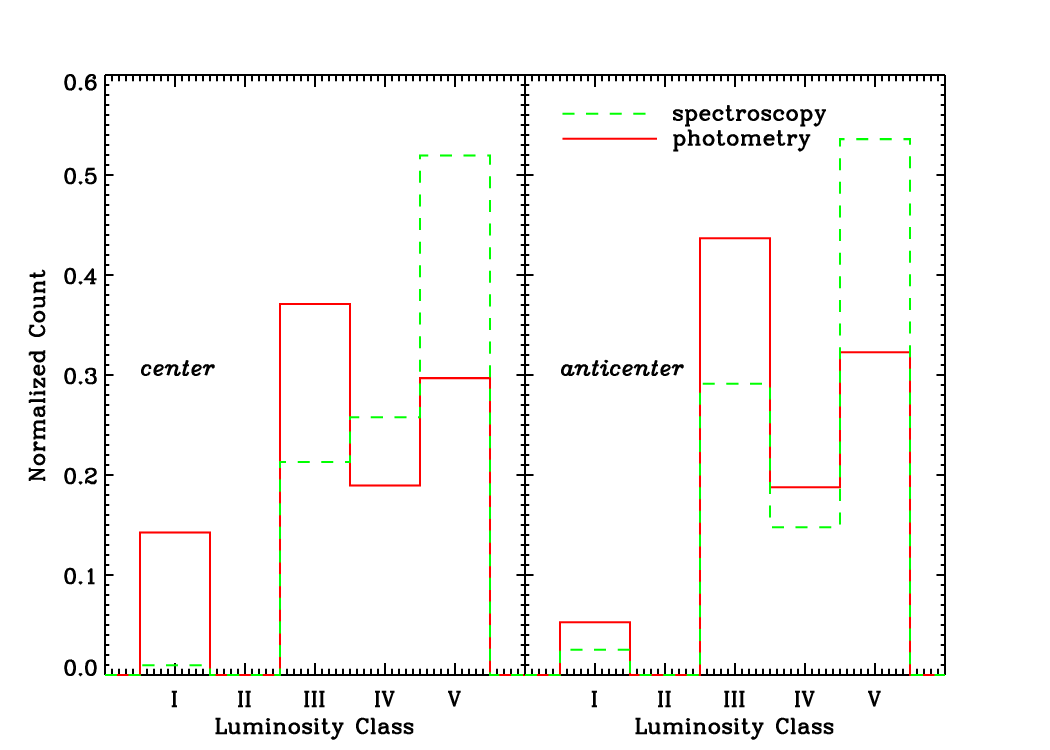}
  \caption{{Comparison of the spectral types and luminosity classes derived with \matisse\ and with the \exodat\ photometric classification criteria.}}
  \label{FighistoComp}
\end{figure*}

The photometric classification that is available in \exodat\ consists of two steps. First, a raw luminosity class was derived from color-magnitude diagrams similar to Fig. \ref{FigCOLMAG}. Then the final estimates of luminosity classes and spectral types were based on spectral energy distribution (SED) fitting with \cite{1998PASP..110..863P} templates. For a full description, we refer to \cite{2009AJ....138..649D}. No preliminary classification was available for stars in the \emph{SRc01} field, hence the \centerf\ simply stands for the \emph{LRc01} field.
In Fig.~\ref{FighistoComp}, we compare the results of the spectral analysis performed by \matisse\ with the photometric classification available in \exodat. The conversion from \teff-\logg\ to spectral type and luminosity class was performed using the \cite{1981Ap&SS..80..353S} tables. Figure~\ref{FighistoComp} illustrates the reasonable agreement found for the spectral types, whereas the luminosity classes histograms show clearly more dwarfs and less giants than estimated by the photometric classification. A discrepancy was expected between the photometric and spectroscopic luminosity classes since the initial photometric estimation of the luminosity class consisted of a simple cut-off in a color-magnitude diagram, applied to prevent the omission of dwarfs during {the \corot} target selection phase. This cut-off could result in a misidentification of dwarfs and giants. For the faint targets, this becomes a more critical issue since the two populations are mixed (Fig.~\ref{FigCOLMAG}). We indeed found a closer agreeement for the brightest targets in our sample.

\section{Discussion \label{SecDis}}

The samples of CoRoT targets for which we determined atmospheric parameters are not representative of the whole fields observed by \corot. As described in Sect.~\ref{SecObs}, we used selection criteria for the multi-fiber observations. As a result, our analysis was focussed primarily on the slowly-rotating population of the FGK stars.
To more accurately characterize the stellar population in the exoplanet fields, we took into account the various biases introduced by our selection. We de biased our samples using density plots of the color-magnitude diagrams presented in Fig.~\ref{FigCOLMAG}. The limitations of this study are constrained by the ranges in color and magnitude covered by our observations (blue boxes in Fig.~\ref{FigCOLMAG}). For every bin of 0.5 in \emph{r'} and 0.5 in \emph{(r'-J)}, we randomly selected stars in our observations so as to fill in the two-dimensional distribution in the same proportions as the \exodat\ one.
The number of stars in the two long run fields was large enough to efficiently apply this procedure. This was not the case for the \emph{SRc01} field to which we did not apply this de-biasing procedure, so the \centerf\ field simply stands for the \emph{LRc01} field. To validate this procedure, we took advantage of the \emph{Besan\c con} Galactic model \citep{2003A&A...409..523R}. We compared each de-biased field with a simulation taking into account the limits in magnitude and assuming the standard extinction law. We found reasonable agreement for the distributions of \teff\ and \logg. 
\begin{figure*}[!h]
  \centering
  \includegraphics[scale=1]{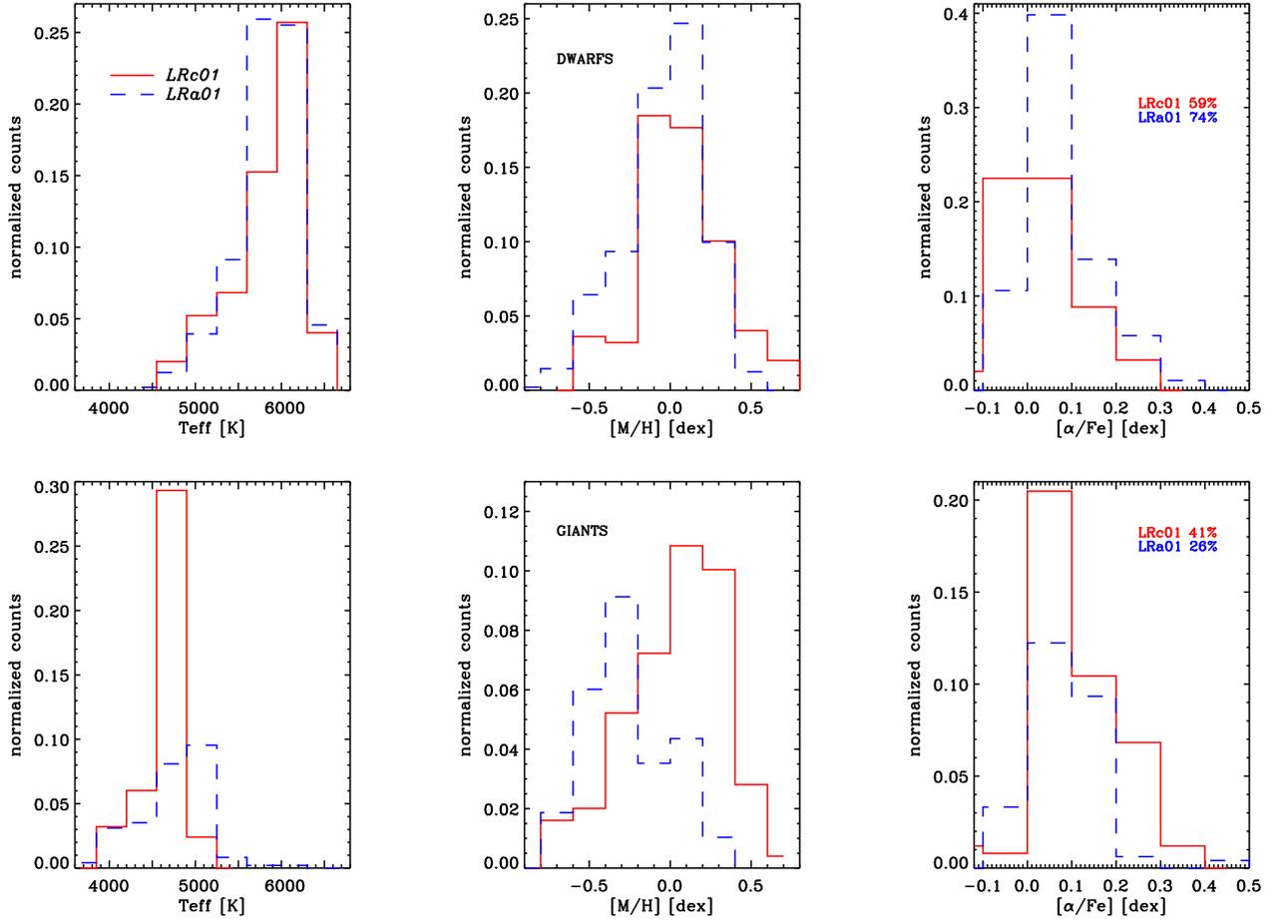}
 \caption{{De-biased distributions of the effective temperature, overall metallicity and $\alpha-$enhancement in the \emph{LRc01} (solid red) and \emph{LRa01} (dashed blue) fields. The top and lower panels show the distributions for the dwarf and giant stars, respectively. The distributions have been normalized to the number of stars in each field. The percentage given in the \alf\ histogram represents the rate of dwarfs (upper panel) and giants (lower panel) in the two fields.}}
  \label{FigHISTOs}
\end{figure*}

To compare the two de~biased {samples, we selected in \emph{LRc01} only stars with $r'\leq15$~mag since this is the magnitude limit of our} observations in the \emph{LRa01} field. For the two fields, we included the stars with an intermediate \logg\ in the giant sample since they are too few to be analyzed separately and they belong mainly to the giant branch in the color-magnitude diagram (Fig.~\ref{FigCOLMAG}).

Figure~\ref{FigHISTOs} shows the resulting distributions of the \teff, \met, and \alf\ derived by \matisse\ for the two fields normalized to the entire de~biased sample. We used the \logg\ {values} from \matisse\ to separate giants from dwarfs as described in the previous section.
The dwarf content is roughly similar in the two directions over these three parameters. The majority of these stars are early G or late F stars with solar metallicity.
The \emph{LRc01} field clearly contains mainly late K giant stars. The other field presents a smoother distribution around K giants, which are slightly hotter.
The histograms of \met\ and \alf\ for the giant stars illustrate the metallicity gradient {generally observed in our galaxy \citep{2009A&A...504...81P}, with the \centerf\ field containing more higher \met\ giant stars than the \anticenter\ one}.
For targets with {$r'\leq15$~mag}, the two fields can be compared: the \emph{LRc01} field is composed of {about 41\% giant and 59\% dwarf stars, whereas} the \emph{LRa01} field is composed of {about 26\% giant and 74\% dwarf stars}. {Adding the objects with $r'\ge15$ spectroscopically observed in the \centerf\ direction,} we estimated the total amount of dwarfs to be 72\% of the stars observed by \corot\ {in this field}.

To estimate the detection rate of exoplanets in the {\emph{LRc01} and \emph{LRa01}} fields, we applied the probability law from \cite{2007ARA&A..45..397U} ($\displaystyle{{\mathcal P} = 0.03\times10^{2.04\times\met}}$ for \met~$\ge 0.0$~dex and $\displaystyle{{\mathcal P}=3\%}$ for \met~$<0.0$~dex) {to the FGK dwarf population with $r'\leq15$}. 
According only to this metallicity criterion, we expected 3.6\% of planets in \emph{LRc01} field, and 3.7\% in the \emph{LRa01} field.
The geometrical probability that a transit occurs is $\displaystyle{{\mathcal P}=R_{*}/a}$, where $a$ is the semi-major axis of the planetary orbit and $R_{*}$ the host star's radius. 
We used the distribution of semi-major axis from the \citet{exoplaneteu:website} website and integrated the transit probability over this distribution.
Combining these probabilities and {applying} them to the two selected fields, we found that the number of expected transiting planets detection is 8$\pm7$ and 9$\pm7$ for the \emph{LRc01} and \emph{LRa01} fields, {respectively, with the error bars simply resulting from the propagation of the error in \met\ (see Sect.~\ref{SecOW})}.

This results is consistent with those derived so far by the \corot/Exoplanet consortium, \textit{i.e.} four transiting planets in the \emph{LRc01} field and three in the LRa01 field.
These estimates must be interpreted with caution because the metallicity probability relation assumed above was derived from planetary detections made using radial velocity techniques whereas the \corot\ space mission bases its detection on planetary transits.

\section{Summary\label{SecConc}}

{Using the \matisse\ algorithm, we have developed a fully automated pipeline to perform a homogeneous spectral analysis of \flames/\giraffe\ spectra covering the Mg~\textsc{I}~b line wavelength range with an intermediate resolving power (HR9B setup)}. 
We have estimated the atmospheric parameters (\teff, \logg), metallicity indices (\met, \alf), the barycentric radial velocity (\vrad), and the projected rotational velocity (\vsini) of more than a thousand of stars located in three of the \corot/Exoplanet fields {towards the Galactic \centerf\ and \anticenter: the \emph{LRa01}, \emph{LRc01}, and \emph{SRc01} fields}.

Comparing these results with the photometric classification available in \exodat, we have found reasonable agreement for the spectral types and a small discrepancy between the luminosity classes. This is due to the difficulty in deriving a luminosity class based only on the broad-band photometry available in \exodat, since this consists of a simple cut-off in the color-magnitude diagrams (see Fig.~\ref{FigCOLMAG}). This limit becomes very uncertain for very faint targets ($r'>14.5$).

From the derived stellar atmospheric parameters, we have compiled the first un-biased reference sample for studying the planet-metallicity relation in the \corot/exoplanet fields.
We found that  the number of planets discovered so far by \corot\ is in agreement with the probability law predicting the number of planets as a function of the stellar metallicity.

In the near future, we will easily be able to implement these efficient procedures to any other \corot\ field. \flames\ observations are currently being performed in other \corot\ pointing directions: the \emph{LRc02} and \emph{LRc03}. The spectral analysis of 1\,000 stars with \matisse\ can be performed within a few minutes using an average computer. This would increase the statistics of available data and help us to understand the typical stellar population {observed by \corot}. Comparing the parameters of the planet-hosting stars with the more general stellar field they belong to, will provide additional constraints on planetary system formation scenarios.

Using the atmospheric parameters with stellar evolution models will lead to the determination of physical parameters for a large number of \corot\ targets.
By adding proper motions to the parameters derived in this work, we will be able to derive kinematical information about these sample stars and help in exploring the age-metallicity and age-kinematics relationships in these fields. Combining this key information with the richness of data provided by the light curves would provide unprecedented insight into these relationships. This will be explored in a forthcoming paper. 

The spectroscopic parameters, the kinematics information derived, and the spectra used in this study will be made available to the community through \exodat\footnote{{http://lamwws.oamp.fr/exodat/}} .

\begin{acknowledgements}
Computations were performed on the ``Mesocentre SIGAMM''
machine, hosted by Observatoire de la C\^ote d'Azur.
We would like to thank J.-C. Bouret for providing the ressources for calculation necessary for this publication.
B. Edvardsson was helpful for the computation of the atmosphere models and B. Plez for the linelist calculations.
We could also use the Elodie3.1 library with the help C. Soubiran.
We  also thank V. Hill for her advice on the delicate process of normalization.

\end{acknowledgements}

\bibliographystyle{aa.bst}
\bibliography{flames}


\begin{table*}[!H]
\caption{{Derived physical parameters for the stars observed in the \emph{LRa01}, \emph{LRc01}, and \emph{SRc01} \corot\ fields. The error bars for the atmospheric parameters are only \emph{internal errors} coming from Fig.~\ref{FigErrIntern}. The \vsini~error is of the order of 10\% of its value. The quality flag (exponent of the value) was set to 1 if the \vsini\ estimate is not affected by noise or a second component in the \ccf; 2 the target is a SB2 and the \vsini\ is related to the main component of the \ccf; 3 the contrast and shape of the \ccf\ are insufficient to assure a proper estimate of its parameters.
{The \sn~of the \flames/\giraffe\ spectrum is listed in the last column of the table. We remind the reader that the coordinates, observing dates and magnitudes are publicly available through \exodat.}\label{TabVrad}}}
\centering
\begin{tabular}{rrrrrrrrrrrrr}
\hline
\hline
\noalign{\smallskip}
 \multicolumn{1}{c}{CoRoT ID} &
 \multicolumn{1}{c}{\vsini} &
 \multicolumn{1}{c}{\vrad} &
 \multicolumn{1}{c}{$\sigma_{V_{\rm rad}}$} &
 \multicolumn{1}{c}{\teff} &
 \multicolumn{1}{c}{$\sigma_{T_{\rm eff}}$} &
 \multicolumn{1}{c}{\logg} &
 \multicolumn{1}{c}{$\sigma_{\log {\it~g} }$} &
 \multicolumn{1}{c}{\met} &
 \multicolumn{1}{c}{$\sigma_{\rm [M/H]}$} &
 \multicolumn{1}{c}{\alf} &
 \multicolumn{1}{c}{$\sigma_{\rm [\alpha/Fe]}$} &
   \multicolumn{1}{c}{\sn} \\
 \multicolumn{1}{c}{} &
 \multicolumn{1}{c}{(\kms)} &
 \multicolumn{1}{c}{(\kms)} &
 \multicolumn{1}{c}{(\kms)} &
 \multicolumn{1}{c}{(K)} &
 \multicolumn{1}{c}{(K)} &
 \multicolumn{1}{c}{(dex)} &
 \multicolumn{1}{c}{(dex)} &
 \multicolumn{1}{c}{(dex)} &
 \multicolumn{1}{c}{(dex)} &
 \multicolumn{1}{c}{(dex)} &
 \multicolumn{1}{c}{(dex)} &
   \multicolumn{1}{c}{} \\

           \noalign{\smallskip}
				\hline
          \noalign{\smallskip}
\object{100603128} & $ 3.8 ^{1}$ &   $ -7.83 $ & $ 0.187 $ & $ 5350 $ & $ 79 $ & $ 4.01 $ & $ 0.13 $ & $ -0.24 $ & $ 0.08 $ & $ 0.06 $ & $ 0.04$ & 9\\
\object{100583300} & $ 17.8^{1}$ & $ -61.10 $ & $ 0.163 $ & $ - $ & $ - $ & $ - $ & $ - $ & $ - $ & $ - $ & $ - $ & $ - $& 23 \\
\object{100567221} & $ 1.4 ^{1}$ &   $ -25.66 $ & $ 0.075 $ & $ 5889 $ & $ 48 $ & $ 4.29 $ & $ 0.08 $ & $ -0.04 $ & $ 0.05 $ & $ 0.02 $ & $ 0.02$&23 \\
\object{100565715} & $ 5.8 ^{1}$ &  $ 92.85 $ & $ 0.246 $ & $ 6445 $ & $ 71 $ & $ 4.64 $ & $ 0.12 $ & $ -0.25 $ & $ 0.07 $ & $ 0.16 $ & $ 0.03$& 12\\
\object{100577769} & $ 8.2 ^{1}$ &  $ -27.26 $ & $ 0.272 $ & $ 5117 $ & $ 76 $ & $ 3.02 $ & $ 0.13 $ & $ -1.02 $ & $ 0.08 $ & $ 0.24 $ & $ 0.04$& 10\\
\object{100588558} & $ 11.3^{1}$ & $ 39.90 $ & $ 0.299 $ & $ - $ & $ - $ & $ - $ & $ - $ & $ - $ & $ - $ & $ - $ & $ - $ & 13\\
\object{100576988} & $ 4.8 ^{1}$ &  $ -26.10 $ & $ 0.118 $ & $ 5963 $ & $ 60 $ & $ 4.41 $ & $ 0.1 $ & $ 0.0 $ & $ 0.06 $ & $ -0.02 $ & $ 0.03$&16 \\
\object{100636016} & $ 2.7 ^{1}$ &  $ 22.55 $ & $ 0.322 $ & $ 5673 $ & $ 80 $ & $ 4.29 $ & $ 0.13 $ & $ -0.88 $ & $ 0.08 $ & $ 0.24 $ & $ 0.04$& 8\\
\object{100637229} & $ 19.4^{1}$ & $ -12.39 $ & $ 0.224 $ & $ - $ & $ - $ & $ - $ & $  $ & $ - $ & $ - $ & $ - $ & $ - $&28 \\
\object{100604545} & $ 10.7^{1}$ & $ 0.37 $ & $ 0.231 $ & $ - $ & $ - $ & $ - $ & $ - $ & $ - $ & $ - $ & $ - $ & $ - $& 10\\
\object{100634976} & $ 3.9 ^{1}$ &  $ -37.27 $ & $ 0.221 $ & $ 6100 $ & $ 73 $ & $ 4.07 $ & $ 0.12 $ & $ -0.35 $ & $ 0.08 $ & $ 0.07 $ & $ 0.04$&11 \\
... & ...&...&...&...&...&...&...&...&...&...&...&...\\
\noalign{\smallskip}  
\hline		      
\end{tabular}	      
\end{table*}

\begin{table*}[!h]
\caption{Derived parameters for the S$^{4}$N sample from \citet{2004A&A...420..183A} (complete table available at the CDS). The first line is the Sun.\label{tabS4N}}
\centering
\begin{tabular}{cccrrccrr}
\hline
\hline
\noalign{\smallskip}  
  \multirow{2}{*}{Hipp. Num.}&
  \multicolumn{1}{c}{\teff} &
  \multicolumn{1}{c}{\logg} &
  \multicolumn{1}{c}{[Fe/H]} &
  \multicolumn{1}{c}{[$\alpha$/Fe]} &
    \multicolumn{1}{c}{\teff\_\matisse} &
  \multicolumn{1}{c}{\logg\_\matisse} &
  \multicolumn{1}{c}{\met\_\matisse} &
  \multicolumn{1}{c}{\alf\_\matisse} \\
  & (K) & (de)] & (dex) &(dex) & (K) & (dex) & (dex) & (dex) \\
\noalign{\smallskip}  
\hline
\noalign{\smallskip}  
           0 &  5777.  &  4.437    &0.000    &0.000    &5724  &  4.227 &   0.036   & 0.029 \\
         171 &  5361. &  4.610 & $-$0.770  & 0.343  & 5376   & 4.359  & $-$0.631  &  0.429\\
         ... & ... & ... & ... & ... & ... & ... & ... & ... \\
\noalign{\smallskip}  
\hline\end{tabular}
\end{table*}

\begin{table*}[!h]
\caption{Derived parameters for the Elodie3.1 sample from \citet{2007astro.ph..3658P} (complete table available at the CDS).\label{tabElo}}
\centering
\begin{tabular}{cccrccrr}
\hline
\hline
\noalign{\smallskip}  
  \multirow{2}{*}{HD identifier}&
  \multicolumn{1}{c}{\teff} &
  \multicolumn{1}{c}{\logg} &
  \multicolumn{1}{c}{[Fe/H]} &
  \multicolumn{1}{c}{\teff\_\matisse} &
  \multicolumn{1}{c}{\logg\_\matisse} &
  \multicolumn{1}{c}{\met\_\matisse} &
  \multicolumn{1}{c}{\alf\_\matisse} \\
  & (K) & (dex) & (dex) & (K) & (dex) & (dex) & (dex) \\
\noalign{\smallskip}  
\hline
\noalign{\smallskip}  
\object{HD000400}   &  6146. &  4.090 & $-$0.280 &  6113 &  4.397 & $-$0.185 & $-$0.075\\
\object{HD001835}   &  5777. &  4.450 &  0.170  & 5883  & 4.804  & 0.197 & $-$0.068\\
 ... & ... & ... & ... & ... & ... & ... & ... \\
\noalign{\smallskip}  
\hline\end{tabular}

\end{table*}

\begin{table*}[!h]
\caption{Derived parameters for the UVES \citep[POP,][]{2003Msngr.114...10B} sample and literature from \citet{2007astro.ph..3658P}.\label{tabPop}}
\centering
\begin{tabular}{cccrccrr}
\hline
\hline
\noalign{\smallskip}  
  \multirow{2}{*}{HD identifier}&
  \multicolumn{1}{c}{\teff} &
  \multicolumn{1}{c}{\logg} &
  \multicolumn{1}{c}{[Fe/H]} &
  \multicolumn{1}{c}{\teff\_\matisse} &
  \multicolumn{1}{c}{\logg\_\matisse} &
  \multicolumn{1}{c}{\met\_\matisse} &
  \multicolumn{1}{c}{\alf\_\matisse} \\
  & (K) & (dex) & (dex) & (K) & (dex) & (dex) & (dex) \\
\noalign{\smallskip}  
\hline
\noalign{\smallskip}  
  \object{HD022049} & 5089.0 & 4.55 & $-$0.10 & 5054 & 4.542 & 0.010 & 0.043\\
  \object{HD022484} & 5989.0 & 4.10 & $-$0.07 & 5916 & 3.891 & $-$0.040 & 0.024\\
  \object{HD030562} & 5860.0 & 4.03 & 0.17 & 5854 & 4.053 & 0.271 & 0.009\\
  \object{HD061421} & 6562.0 & 4.08 & $-$0.02 & 6366 & 3.990 & 0.067 & $-$0.085\\
  \object{HD076932} & 5850.0 & 4.05 & $-$0.93 & 5600 & 3.593 & $-$1.052 & 0.391\\
  \object{HD115383} & 5979.0 & 4.14 & 0.08 & 6063 & 4.470 & 0.229 & $-$0.047\\
  \object{HD128167} & 6782.0 & 4.32 & $-$0.40 & 6658 & 4.543 & $-$0.214 & $-$0.022\\
\noalign{\smallskip}  
\hline\end{tabular}
\end{table*}

\begin{table*}[!h]
\caption{Derived parameters for the \citet{2009A&A...493..309S} sample (complete table available at the CDS).\label{tabSan}}
\centering
\begin{tabular}{lccrccrr}
\hline
\hline
\noalign{\smallskip}  
  \multirow{2}{*}{Identifier}&
  \multicolumn{1}{c}{\teff} &
  \multicolumn{1}{c}{\logg} &
  \multicolumn{1}{c}{[Fe/H]} &
  \multicolumn{1}{c}{\teff\_\matisse} &
  \multicolumn{1}{c}{\logg\_\matisse} &
  \multicolumn{1}{c}{\met\_\matisse} &
  \multicolumn{1}{c}{\alf\_\matisse} \\
  & (K) & (dex) & (dex) & (K) & (dex) & (dex) & (dex) \\
\noalign{\smallskip}  
\hline
\noalign{\smallskip} 
\object{IC2714No110} &     5017.&   2.850 &  0.010 & 5028  & 2.911&  $-$0.033 &  0.047\\
\object{IC2714No87} &      5029. &  2.620&  $-$0.060 & 5009 &   2.889 & $-$0.181 & $-$0.024\\
 ... & ... & ... & ... & ... & ... & ... & ... \\
\noalign{\smallskip}  
\hline\end{tabular}
\end{table*}

\end{document}